\documentclass[review, fleqn, 3p]{elsarticle}

\usepackage{picinpar}
\usepackage{url}
\usepackage{flushend}
\usepackage{colortbl}
\usepackage{soul}
\usepackage{multirow}
\usepackage{pifont}
\usepackage{color}
\usepackage{alltt}
\usepackage{enumerate}
\usepackage{siunitx}
\usepackage{breakurl}
\usepackage{epstopdf}
\usepackage{pbox}

\usepackage{booktabs}
\usepackage[T1]{fontenc}
\usepackage{setspace}

\usepackage[tight]{subfigure}
\usepackage{amsmath,amssymb,amscd,graphicx, color}

\allowdisplaybreaks

\newcommand{\abs}[1]{\left| #1 \right|}

\newcommand*{\red}{\textcolor{black}}

\begin{document}


\begin{frontmatter}

\title{Impedance-based Capacity Estimation for Lithium-Ion Batteries Using Generative Adversarial Network}

\author[mymainaddress]{Seongyoon Kim\fnref{cor1}}
\author[mymainaddress]{Yun Young Choi\fnref{cor1}}
\author[mymainaddress]{Jung-Il Choi\corref{cor2}}\ead{jic@yonsei.ac.kr}
\address[mymainaddress]{School of Mathematics and Computing (Computational Science and Engineering), Yonsei University, Seoul 03722, Republic of Korea}
\fntext[cor1]{These authors contributed equally to this work and should be considered co-first authors.}
\cortext[cor2]{Corresponding author.}

\begin{abstract}

This paper proposes a fully unsupervised methodology for the reliable extraction of latent variables representing the characteristics of lithium-ion batteries (LIBs) from electrochemical impedance spectroscopy (EIS) data using information maximizing generative adversarial networks.
Meaningful representations can be obtained from EIS data even when measured with direct current and without relaxation, which are difficult to express when using circuit models.
The extracted latent variables were investigated as capacity degradation progressed and were used to estimate the discharge capacity of the batteries by employing Gaussian process regression.
The proposed method was validated under various conditions of EIS data during charging and discharging.
The results indicate that the proposed model provides more robust capacity estimations than the direct capacity estimations obtained from EIS.
We demonstrate that the latent variables extracted from the EIS data measured with direct current and without relaxation reliably represent the degradation characteristics of LIBs.

\end{abstract}


\end{frontmatter}

\section{Introduction}
Lithium-ion batteries (LIBs) are prominent energy storage solutions that have been implemented in various applications.
Their high energy density, long lifespan, and low self-discharge make these batteries suitable for applications in electric vehicles (EVs) and energy storage systems~\cite{armand2008building, dunn2011electrical}.
However, issues pertaining to the reliability and safety of LIBs still remain challenging to address~\cite{tarascon2011issues, vetter2005ageing, jaguemont2016comprehensive, gandoman2019concept}; therefore, monitoring the state of health (SOH) is essential for ensuring the safe handling, maintenance, and replacement of LIBs~\cite{zhang2011review, rahimi2013battery, waag2014critical}.
The SOH is defined as the ratio of the usable capacity in the current cycle to the initial or rated capacity specified by the manufacturer.
The SOH cannot be measured directly; instead, it must be inferred from the internal characteristics of the battery, which are estimated based on constant current (CC)-constant voltage (CV) charging, current pulses, or electrochemical impedance spectroscopy (EIS)~\cite{yang2018online, xiong2018towards, li2020digital, hu2020enhanced}.


Recent studies have revealed that battery impedance is strongly dependent on the SOH~\cite{xiong2017systematic, pastor2017comparison, zhang2020identifying, zhang2021electrochemical}.
EIS is a common approach used to determine battery impedance; in this approach, a sinusoidal perturbation is applied to the battery current~\cite{lasia2014definition, meddings2020application}.
To obtain the relationship between EIS and SOH, equivalent circuit models (ECMs) are widely used, and the measured impedance is fitted to nonlinear parameters~\cite{chen2006accurate, vyroubal2018equivalent, zheng2021simplification}.
Most frequently, EIS measurements are presented in the Nyquist plane, and data regression methods are applied to obtain the ECM parameters~\cite{stroe2014diagnosis, lai2019comparative, sihvo2020soh}.
%
%
Although EIS analysis is a standard tool used for characterizing LIBs, EIS is highly sensitive to the test conditions~\cite{andre2011characterization, lyu2019situ}.
Therefore, a long rest time is typically needed to ensure that the battery is in electrochemical equilibrium before conducting the EIS measurements~\cite{westerhoff2016analysis}.
As such, estimating the state of LIBs by using EIS requires stationary and stable conditions, which can increase the experimental costs.
However, if these specific experimental conditions are not adopted, fluctuations in the low-frequency region of the EIS measurements may occur, making the interpretation difficult~\cite{sun2014resolving}.
Moreover, model-based approaches can be inadequate for representing all the characteristics of LIBs under various operating conditions.
Hence, we focused on data-driven approaches for reliable diagnoses through the robust characterization of EIS data involving low-frequency fluctuations.

In recent years, data-driven approaches have been considered to be promising because they neither require prior knowledge of the battery systems nor depend on specific systems~\cite{nuhic2013health, richardson2017gaussian, gou2020state, mawonou2021state}.
In particular, approaches based on neural networks directly approximate high-dimensional nonlinear functions from raw data and achieve high prediction accuracy in solving complex problems.
Specifically, {this study focuses on unsupervised learning~\cite{li2019big, fasahat2020state, zhang2020implementation}, a data-driven method which extracts values from specific unlabeled data.
The unsupervised model can learn meaningful parameters related to the SOH from the EIS data, as in the ECM}.
Improvements in unsupervised learning architectures, such as generative adversarial networks (GANs)~\cite{goodfellow2014generative} and variational autoencoders (VAEs)~\cite{kingma2013auto}, have gained attention for learning probability distributions over complex data.
Both the GAN and VAE exhibit satisfactory data generation results; however, the GAN is unable to extract features from data, while the VAE is generally known to incur a loss of fidelity, even though it employs both an encoder and a decoder.
The information-maximizing GAN (InfoGAN) was recently developed to extract interpretable and meaningful representations by introducing a variational lower bound, which maximizes the mutual information between latent variables and the observations~\cite{chen2016infogan}.
Using InfoGAN, meaningful latent variables can be extracted in a fully unsupervised manner from EIS measurements that are difficult to express using circuit models.
There are a few studies which indicate the use of InfoGAN to diagnose a failure of engines or bearing~\cite{zheng2019generative, wu2020ss}.
However, to the best of our knowledge, this study is the first attempt to obtain the latent variables from the impedance of LIBs by applying InfoGAN.

This paper presents an EIS-based InfoGAN (EISGAN) to extract latent variables that can reliably represent battery characteristics from EIS data with low-frequency fluctuations.
\red{
Typically, a circuit model is used to analyze the EIS data obtained from laboratory experiments.
However, in practical conditions with direct current (DC) and without relaxation, the EIS data can be observed as having various shapes, which are difficult to express using limited circuit model parameters.
Therefore, we have applied the InfoGAN, which can extract meaningful parameters related to the SOH from the EIS data as in ECM, to interpret the various shapes.
}
Through analyses of EIS measurements under various conditions according to relaxation and DC, changes in the latent variables over cycles were investigated, as the capacity degradation progressed.
The extracted latent variables were used to estimate the discharge capacity of the batteries by using the Gaussian process regression (GPR).
The simulation results indicate more robust estimations than directly predicting the capacity through machine learning based on the EIS.
Notably, the novelty of this study lies in the reliable estimation of the latent variables involving degradation characteristics of LIBs from the EIS data, with DC and without relaxation, in a fully unsupervised model-free manner.
We demonstrate reliable estimations of the latent variables for the EIS of LIBs in a model-free manner and indicate that EIS data with DC and without relaxation involve the degradation characteristics of LIBs.

The remainder of this paper is organized as follows.
Section~\ref{sec2:method} presents the proposed EISGAN architecture and the procedure for capacity estimation.
Descriptions regarding the data and detailed implementations are presented in Section~\ref{sec3:precedure}, and Section~\ref{sec4:results} presents the evaluation of the EISGAN and capacity predictions.
Lastly, Section~\ref{sec5:conclusion} presents the major conclusions drawn from this study and outlines the scope of future research.


\section{Methodology}\label{sec2:method}

\subsection{EISGAN for latent estimation}\label{sec:EISGAN}

\begin{figure}[t]
    \centering
    \includegraphics[width=.8\columnwidth]{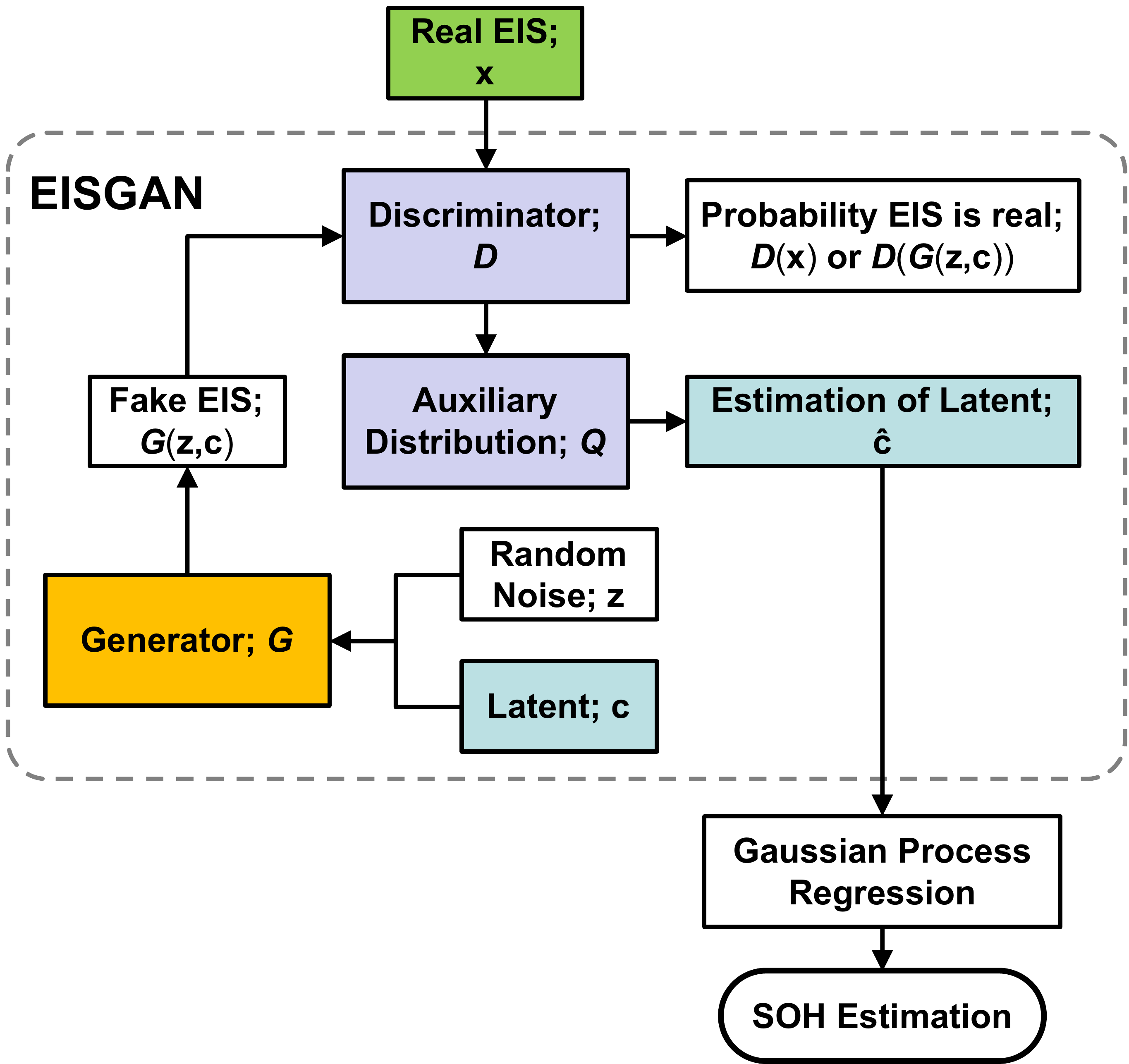}
    \caption{Network structure of EISGAN and procedure of SOH estimation using GPR.}
    \label{fig:EISGAN_Structure}
\end{figure}

The structure of the proposed EISGAN is illustrated in Fig.~\ref{fig:EISGAN_Structure}.
The EISGAN is constructed based on the InfoGAN~\cite{chen2016infogan}, which can extract interpretable and meaningful latent variables by introducing a variational lower bound.
Essentially, InfoGAN is a generative model, and the goal is to learn the generator distribution $P_G(\mathbf{x})$ that matches the actual data distribution $P_{data}(\mathbf{x})$.
Instead of explicitly assigning probability for all $\mathbf{x}$ in the data distribution, InfoGAN learns a network $G$ generating data $G(\mathbf{c},\mathbf{z})$ from meaningful latent variables $\mathbf{c}\sim P_{latent}(\mathbf{c})$ and noise variables $\mathbf{z}\sim P_{noise}(\mathbf{z})$.
The adversarial discriminator network $D$ is also trained against the generator and aims to distinguish between the true data distribution $P_{data}$ and the sample from the generator's distribution, $P_G$.
The minimax loss function of the generator and the discriminator is formally defined as
\begin{align}
    \min_G\max_D V(D,G)&=\mathbb{E}_{\mathbf{c}\sim P_{latent},\mathbf{z}\sim P_{noise}} [\log (1-D(G(\mathbf{c},\mathbf{z})))] + \mathbb{E}_{\mathbf{x}\sim P_{data}}[\log D(\mathbf{x})],
\end{align}
where $\mathbb{E}_{\mathbf{p}\sim P}$ denotes an expectation for some variables, $\mathbf{p}$, which follow a distribution, $P$.

In the InfoGAN, the information in the latent variables, $\mathbf{c}$, should not be lost in the generation process.
Essentially, the mutual information, $I(\mathbf{c} ; G(\mathbf{c},\mathbf{z}))$, which measures the amount of information about the latent variables $\mathbf{c}$ learned from the knowledge of generated data, $G(\mathbf{c},\mathbf{z})$, is required to be maximized as follows.
\begin{align}
    I(\mathbf{c} ; G(\mathbf{c},\mathbf{z})) = H(\mathbf{c}) - H(\mathbf{c} | G(\mathbf{c},\mathbf{z})),
\end{align}
where $H$ stands for entropy.
However, the mutual information is difficult to maximize directly since it requires access to the posterior distribution, $P(\mathbf{c} | \mathbf{x})$.
Therefore, an auxiliary distribution network, $Q$, is introduced to approximate the posterior distribution of latent variables $\mathbf{c}$ by obtaining a variational lower bound $L_I(G,Q)$ of the mutual information of $\mathbf{c}$ and $G(\mathbf{c},\mathbf{z})$, which can be defined as:
\begin{align}
    L_I (G,Q) &\equiv H(\mathbf{c}) + \mathbb{E}_{\mathbf{c}\sim P_{latent},\mathbf{x}\sim P_{G}}[\log Q(\mathbf{c}|\mathbf{x})] \\
    &\le I(\mathbf{c} ; G(\mathbf{c},\mathbf{z})),
\end{align}
because
\begin{align}
    H(\mathbf{c} | G(\mathbf{c},\mathbf{z}))
    &= -\mathbb{E}_{\mathbf{x}\sim P_{G}}[\mathbb{E}_{\mathbf{c}\sim P(\mathbf{c}|\mathbf{x})}[\log P(\mathbf{c}|\mathbf{x})]] \\
    &\le -\mathbb{E}_{\mathbf{x}\sim P_{G}}[\mathbb{E}_{\mathbf{c}\sim P(\mathbf{c}|\mathbf{x})}[\log Q(\mathbf{c}|\mathbf{x})]] \\
    &= -\mathbb{E}_{\mathbf{c}\sim P_{latent},\mathbf{x}\sim P_{G}}[\log Q(\mathbf{c}|\mathbf{x})]
    .
\end{align}
This technique is known as variational information maximization~\cite{agakov2004algorithm}.

For the EIS measurements $\mathbf{x}$, the EISGAN is trained to solve the minimax loss function $V(D,G,Q)$, which is defined as
\begin{align}\label{eq:minimax_loss}
    \min_{G,Q}\max_D V_I(D,G,Q)&=V(D,G)
    - \lambda L_I (G,Q),
\end{align}
where $\lambda$ is a predefined regularization factor.
In \eqref{eq:minimax_loss}, the discriminator $D$ is trained to maximize the probability of assigning the correct label to both the true and generated EIS, and the generator $G$ is trained to minimize the probability that $D$ makes the right decision regarding the generated EIS.
Besides, the auxiliary distribution network $Q$ is trained to maximize the variational lower bound $L_I(G,Q)$.
Note that $H(\mathbf{c})$ is independent of the network, {$Q$}; hence, it was not considered in the training.
By maximizing the lower bound, which leads to the maximization of the mutual information of $\mathbf{c}$ and $G(\mathbf{c},\mathbf{z})$, the information of the latent variables will be retained during the generation process.
A mathematical treatment is presented in \cite{chen2016infogan}.
Finally, using the auxiliary distribution $Q$, the EISGAN extracts meaningful latent variables $\mathbf{c}^*$ from the new EIS data $\mathbf{x}^*$, as follows:
\begin{align}\label{eq:latent}
    \mathbf{c}^*=Q(D(\mathbf{x}^*))
\end{align}

We used a convolution neural network (CNN) to construct the networks $D$, $G$, and $Q$.
In this study, one-dimensional (1D) convolution layers were applied to the EIS frequency domain.
The input EIS sample was assumed to be $\mathbf{x}_r=[x_{1,r},x_{2,r},\ldots,x_{T,r}]$, where $r=1,2,\ldots,R$ and $T$ are the channel index (here, $R=2$ for real and imaginary values of EIS data) and the length of the frequency sequence, respectively.
The convolution operation $\otimes$ in the convolution layer uses a $k$-th kernel $\mathbf{w}_k\in \mathbb{R}^{D\times1}$ with a kernel size $D$.
The $t$-th element of the $k$-th channel output can be expressed as
\begin{align}\label{eq:convolution}
    \left(\mathbf{x}\otimes\mathbf{w}\right)_{t,k} &= \sum_{r=1}^R\left[x_{t,r}, x_{t+1,r},\ldots,x_{t+D-1,r}\right]\mathbf{w}_k + b_k,
\end{align}
where $b_k$ denotes the bias of the $k$-th kernel.
Subsequently, the following convolution layer receives the outputs from the previous convolution layer.
The convolution layers are connected with nonlinear activation functions.
It should be noted that $Q$ and $D$ share all the convolution layers except for the last layer, in order to reduce the computational cost, in accordance with \cite{chen2016infogan}.

\subsection{GPR for capacity estimation}

The GPR is used to estimate capacity by employing the latent variables extracted from the EISGAN.
GPR is a non-parametric regression method that can approximate the nonlinear relationship between inputs and outputs and obtain the predictive distribution~\cite{rasmussen2003gaussian, zhang2020identifying}.
In GPR, the measurements $\mathbf{y}=[y_1,y_2,\ldots,y_n]^T$ can be modeled by inputs $\mathbf{C}=[\mathbf{c}_1,\mathbf{c}_2,\ldots,\mathbf{c}_n]^T$, as follows:
\begin{align}\label{eq:GP_function}
    y_i = f(\mathbf{c}_i) + \epsilon_i,~i=1,2,\ldots,n,
\end{align}
where $f$ is the implicit function approximated via GPR, and $\epsilon_i\sim N(0, \sigma_n^2)$ is the independent and identically distributed Gaussian noise.
In this study, the inputs $\mathbf{c}_i$ are latent variables extracted from EIS measurements by EISGAN at the $i$-th cycle, and the output $y_i$ is the capacity at the corresponding cycle.
The outputs of GPR, $\mathbf{f}=[f(\mathbf{c}_1,\mathbf{c}_2,\ldots,\mathbf{c}_n)]^T$, are modeled as a Gaussian random field $\mathbf{f}\sim N(0,\mathbf{K})$, where $K_{ij}=k(\mathbf{c}_i,\mathbf{c}_j)$ is a covariance kernel matrix that measures the closeness between the points $\mathbf{c}_i$ and $\mathbf{c}_j$.
We used a squared-exponential kernel function (also called a radial basis function) expressed as
\begin{align}\label{eq:kernel}
    k(\mathbf{c}_i,\mathbf{c}_j)=\sigma_f^2\exp\left(\frac{-(\mathbf{c}_i-\mathbf{c}_j)\cdot(\mathbf{c}_i-\mathbf{c}_j)}{2l^2}\right),
\end{align}
where $\cdot$ represents the inner product, $\sigma_f$ determines the amplitude of the kernel function, and $l$ is the length scale of the distance measure.
The hyper-parameters $\Theta=[\sigma_n, \sigma_f,l]$ are then optimized to maximize the log-likelihood $\mathcal{L}$, expressed as
\begin{align}
    \mathcal{L}&=-\frac{1}{2}\log\left(\det\left(\mathbf{K}(\mathbf{C},\mathbf{C})+\sigma_n^2I\right)\right) - \frac{1}{2}\mathbf{y}^T\left(\mathbf{K}(\mathbf{C},\mathbf{C})+\sigma_n^2I\right)^{-1}\mathbf{y} - \frac{n}{2}\log 2\pi,
\end{align}
where $\det$ represents the determinant of a matrix, and $I$ is a unit matrix.
After the GPR is trained, a new data set $\mathbf{c}^*$ that follows the same Gaussian distribution as the training set $\mathbf{C}$, the estimated posterior distribution of $y^*$ can be derived based on Bayes' theorem.
The joint prior distribution of $\mathbf{y}$ and predicted value $y^*$ can be expressed as
\begin{align}\label{eq:prior}
    \begin{bmatrix}
    \mathbf{y} \\ y^*
    \end{bmatrix}
    \sim N\left(0,
    \begin{bmatrix}
    \mathbf{K}(\mathbf{C},\mathbf{C})+\sigma_n^2I & \mathbf{K}(\mathbf{C},\mathbf{c}^*)\\
    \mathbf{K}(\mathbf{c}^*,\mathbf{C}) & \mathbf{K}(\mathbf{c}^*,\mathbf{c}^*)
    \end{bmatrix}
    \right).
\end{align}
Finally, the predicted distribution of the capacity $\mathbf{y}^*$ conditioned on the training set can be obtained as
\begin{align}
    y^*|\mathbf{C}, \mathbf{y}, \mathbf{c}^*\sim N(\overline{y}^*, \sigma_{y^*}^2),
\end{align}
where
\begin{align}
    \overline{y}^* &= \mathbf{K}(\mathbf{c}^*, \mathbf{C})\left(\mathbf{K}(\mathbf{C}, \mathbf{C})+\sigma_n^2I\right)^{-1}\mathbf{y} \label{eq:GP_mean}\\
    \sigma_{y^*}^2 &= \mathbf{K}(\mathbf{c}^*, \mathbf{c}^*)- \mathbf{K}(\mathbf{c}^*, \mathbf{C})\left(\mathbf{K}(\mathbf{C}, \mathbf{C})+\sigma_n^2I\right)^{-1}\mathbf{K}(\mathbf{C}, \mathbf{c}^*). \label{eq:GP_var}
\end{align}


\section{Experimental data and implementation}\label{sec3:precedure}

\subsection{Experimental data}

\begin{figure}[t]
    \centering
    \includegraphics[width=.8\columnwidth]{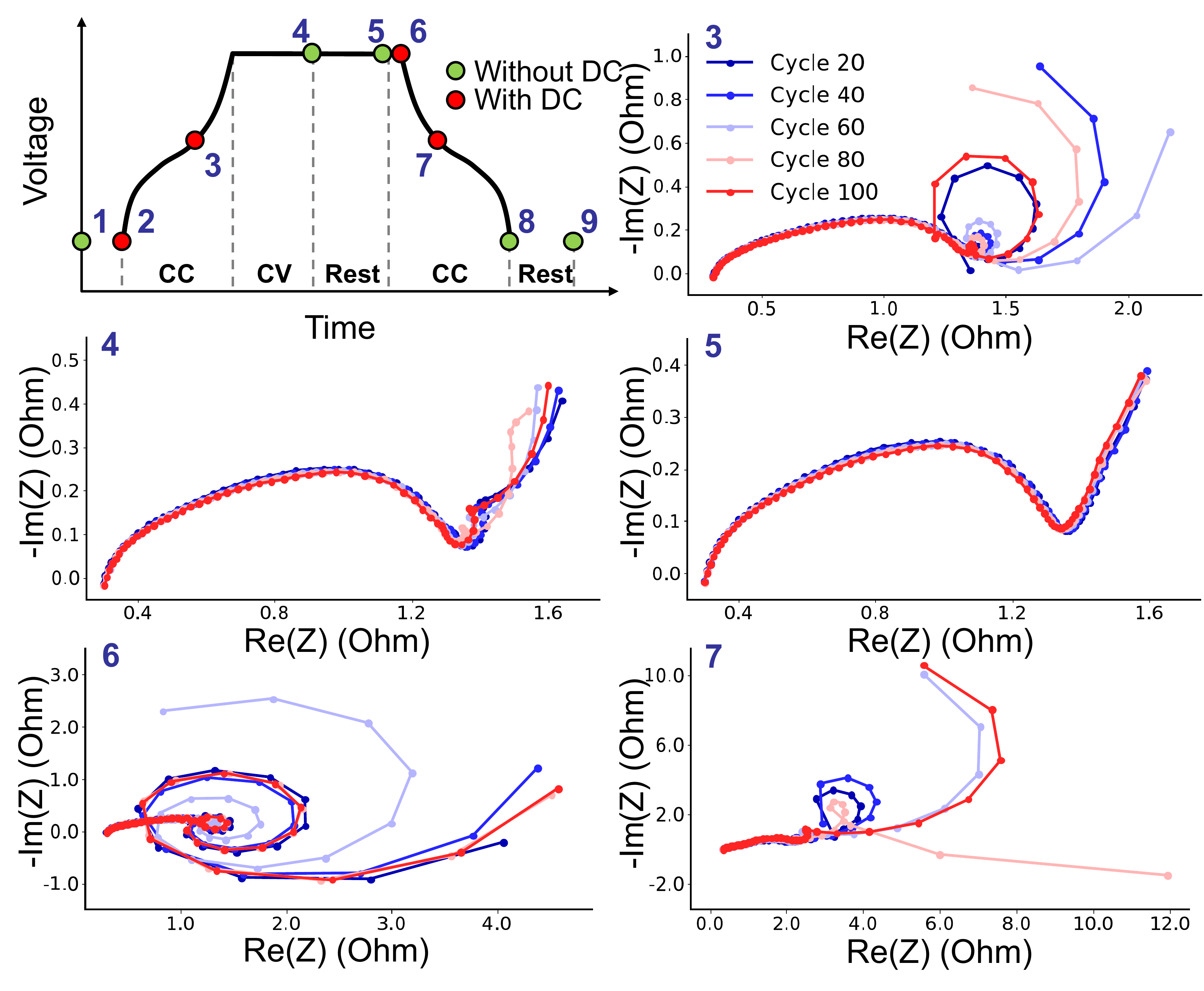}
    \caption{EIS data {of 25C06} collected at different stages during CC-CV charging and CC discharging over 20, 40, 60, 80, and 100 cycles. Only stages with a non-zero SOC are considered.}
    \label{fig:EIS_Stages}
\end{figure}

The proposed EISGAN is validated using EIS data published in \cite{zhang2020identifying}, where commercially available 45-mAh Eunicell LR2032 lithium-ion coin cells were cycled with $1C$ CC-CV charging and $2C$ CC discharging.
EIS measurements in the frequency range of 0.02--20 kHz with an excitation current of 5 mA were performed at nine different stages according to the state of charge (SOC) with and without relaxation and DC.
Eight cells cycled in climate chambers set to 25\textdegree{}C (named as 25C01--25C08), containing EIS and capacity data according to cycles, were used in this study.
\red{
The descriptions of each stage are summarized in Table~\ref{tab:description}.
}
The changes in the EIS measurements {of 25C06} with respect to the cycles in stages 3, 4, 5, 6, and 7 are shown in Fig.~\ref{fig:EIS_Stages}, where Im(Z) and Re(Z) are the imaginary and real parts of the impedance Z, respectively.
It should be noted that the stages were selected to represent conditions according to the presence or absence of relaxation and DC.
Stage 5 is an ideal condition with a 15-min resting period after fully charge.
In stage 4, immediately after charging, slight fluctuations were observed in the low-frequency region.
On the other hand, the EIS data were measured during the $1C$ CC charging in stage 3 (after 20 min of charging) and the $2C$ CC discharging in stages 6 (start of discharging) and 7 (after 10 min of discharging).
Severe fluctuations in the low-frequency region were observed; consequently, it was difficult to fit the ECM with the EIS data.
Among the stages with DC, stage 6 represents the condition after relaxation, while the remaining stages do not involve relaxation.
In addition, the EIS measurements for the other cells are illustrated in Fig.~\ref{fig:EIS_All} in the Appendix.

\begin{table}[t]
\centering
\caption{Descriptions and presence ($\checkmark$) or absence (-) of relaxation and DC for each stage.}\label{tab:description}
\begin{tabular}{llll}
\toprule
Stage & Description                          & Resting & DC \\
\midrule
1     & Before charging                      & $\checkmark$       & -  \\
2     & Start charging                       & $\checkmark$       & $\checkmark$  \\
3     & After 20-min charging                & -       & $\checkmark$  \\
4     & After charging and before resting    & -       & -  \\
5     & After 15-min rest                    & $\checkmark$       & -  \\
6     & Start discharging                    & $\checkmark$       & $\checkmark$  \\
7     & After 10-min discharging             & -       & $\checkmark$  \\
8     & After discharging and before resting & -       & -  \\
9     & After 15-min rest                    & $\checkmark$       & -  \\
\bottomrule
\end{tabular}
\end{table}

\subsection{Model implementation}\label{sec:implemenddetail}

For validating the performance of the model, the EISGANs were separately configured for each of the five stages (i.e., 3, 4, 5, 6, and 7).
Among the eight batteries, 25C01--25C04 were set as the training set, whereas 25C05--25C08 were set as the testing set.
Notably, only data from batteries 25C04, 25C06, 25C07, and 25C08 exist in stage 7; hence, the first two were set as the training set, and the remaining were set as the testing set.
The procedure of the proposed capacity estimation method can be categorized into two processes involving the EISGAN and GPR:

\begin{enumerate}
    \item EISGAN:

    \begin{enumerate}
        \item EISGAN was constructed using 1D convolution layers for $D$, $G$, and $Q$ networks, including $R=2$ channels representing the real and imaginary parts of the EIS data and $T=60$ points on the frequency domain.
        \red{
        Therefore, an input entering the EISGAN for the EIS curve $Z$ under frequency $f_1,f_2,\ldots,f_T$ can be presented as follows:
        \begin{align}
            \mathbf{x}=
            \begin{bmatrix}
                Re(Z(f_1)) & Re(Z(f_2)) & \cdots & Re(Z(f_T)) \\
                Im(Z(f_1)) & Im(Z(f_2)) & \cdots & Im(Z(f_T))
            \end{bmatrix}
            \in \mathbb R^{R\times T}.
        \end{align}
        }
        Leaky rectified linear unit activation functions were applied to every layer, which is expressed as
        \begin{align}
            \phi(\mathbf{x})=\max(\alpha\mathbf{x}, \mathbf{x}),
        \end{align}
        where the negative slope $\alpha$ was set to $0.01$.
        The number of latent variables was set to nine, similar to the number of estimable parameters of the ECM.
        Note that the circuit used for the ECMs, applied to analyze the EIS data, consists of approximately 7--11 parameters~\cite{steinhauer2017investigation, farmann2018comparative}.

        \item The EISGAN was trained to solve \eqref{eq:minimax_loss} using the EIS data of the training cells.
        During model training, the adaptive moment estimation with projection (AdamP) optimizer~\cite{heo2020adamp} was used, which enables stable and robust learning by suppressing the weight norm growth.
        The learning rates of $D$, $G$, and $Q$ were set to 0.0004, 0.0001, and 0.0001, respectively, and a regularization factor $\lambda=0.1$ was adopted.

        \item By using the auxiliary distribution network $Q$ of the completely trained EISGAN, latent variables were obtained using EIS measurements $\mathbf x$ of each cycle, as in \eqref{eq:latent}.
        The latent variables obtained from the training \red{(25C01--25C04)} and testing \red{(25C05--25C08)} cells were named $\mathbf{C}_{train}$ and $\mathbf{C}_{test}$, respectively\red{, and can be expressed as follows:
        \begin{align}
            \mathbf{C}_{train}=
            \begin{bmatrix}
                \mathbf{C}^{(1)} \\ \mathbf{C}^{(2)} \\ \mathbf{C}^{(3)} \\ \mathbf{C}^{(4)}
            \end{bmatrix}~\text{and}~
            \mathbf{C}_{test}=
            \begin{bmatrix}
                \mathbf{C}^{(5)} \\ \mathbf{C}^{(6)} \\ \mathbf{C}^{(7)} \\ \mathbf{C}^{(8)}
            \end{bmatrix},
        \end{align}
        where $\mathbf{C}^{(j)}=[\mathbf{c}_1,\mathbf{c}_2,\ldots,\mathbf{c}_{n_j}]^T$.
        The corresponding capacity values were named $\mathbf{y}_{train}$ and $\mathbf{y}_{test}$, respectively.
        }

    \end{enumerate}

    \item GPR:

    \begin{enumerate}
        \item After obtaining the latent variables through the EISGAN, the latent variables, $\mathbf{C}_{train}$, of the training batteries were set as inputs and the corresponding capacity $\mathbf{y}_{train}$ was set as the output corresponding to the cycles.
        \red{
        The GPR was constructed to model $\mathbf{y}_{train}$ from $\mathbf{C}_{train}$, as shown in \eqref{eq:GP_function}.
        Initially, the covariance kernel matrix $\mathbf{K}$ was obtained from $\mathbf{C}_{train}$ using the squared-exponential kernel expressed in \eqref{eq:kernel}.
        }
        The GPR was then trained by optimizing the hyper-parameters, $\Theta=[\sigma_n, \sigma_f,l]$, to maximize the log-likelihood $\mathcal{L}$ in (7).
        Note that the GPR was implemented using scikit-learn~\cite{scikit-learn}.

        \item
        \red{
        To estimate the predictive distribution of capacity using the latent variables $\mathbf{C}_{test}$ of the testing batteries, the joint prior distribution of $\mathbf{y}_{train}$ and $\mathbf{y}_{test}$ was constructed, as shown in \eqref{eq:prior}.
        }
        Finally, the predictive distribution of $\mathbf{y}_{test}$ conditioned on the training set was obtained, where the mean and confidence interval of the capacity prediction were calculated using \eqref{eq:GP_mean} and \eqref{eq:GP_var}.

    \end{enumerate}

\end{enumerate}

To evaluate the performance of the proposed method, the mean absolute error (MAE), root mean squared error (RMSE), and coefficient of determination ($R^2$) of the capacity were calculated as follows.
\begin{align}
    \text{MAE}\text{~(mAh)} &= \frac{1}{n}\sum_{i=1}^n \abs{y_{i} - \hat y_{i}}\\
    \text{RMSE}\text{~(mAh)} &= \sqrt{\frac{1}{n}\sum_{i=1}^n (y_{i} - \hat y_{i})^2}\\
    R^2 &= 1-\frac{\sum_{i=1}^n (y_{i} - \hat y_{i})^2}{\sum_{i=1}^n(y_{i} - \overline y)^2},
\end{align}
where $y_i$ and $\hat y_{i}$ denote the true and estimated capacities at the $i$-th cycle, respectively, and $\overline y=\frac{1}{n}\sum_{i=1}^n y_i$.


\section{Results and discussion}\label{sec4:results}

\subsection{Latent variables from EIS}

\begin{figure}[t]
    \centering
    \includegraphics[width=.8\columnwidth]{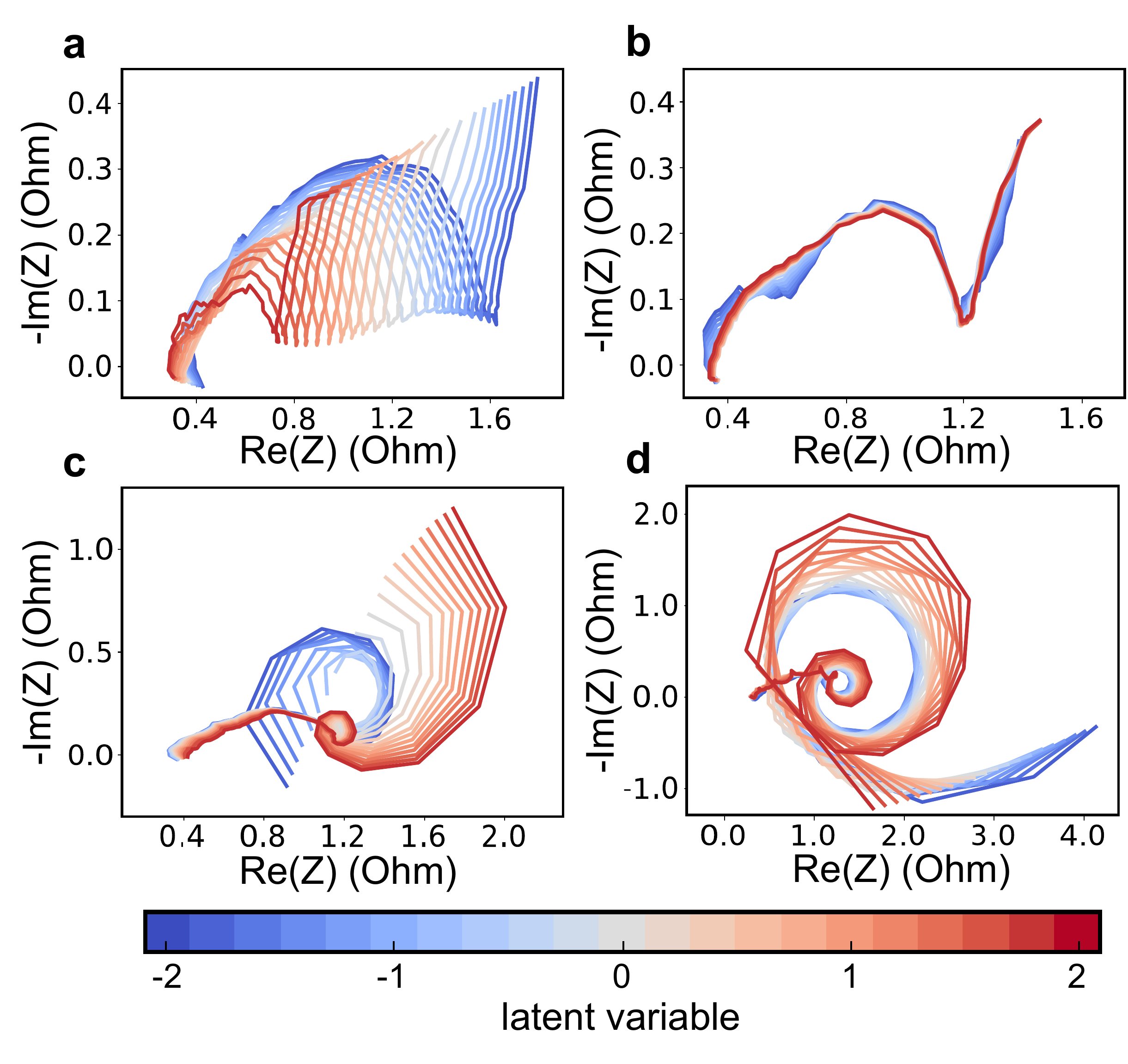}
    \caption{Generated EIS data with latent variables (a) $c_1$ and (b) $c_2$ for stage 5, (c) $c_1$ for stage 3, and (d) $c_1$ for stage 6 varying from $-2$ to $2$ when using EISGAN.}
    \label{fig:GeneratedEIS}
\end{figure}

Generating reliable EIS data is a crucial factor in obtaining credible latent variables.
Therefore, using the generator $G$ of the EISGAN trained for each stage, changes in the generated EIS $\mathbf{x}_{g}$ were investigated when each latent variable changed from $-2$ to $2$, as follows:
\begin{align}
    \mathbf{x}_{g}=G(\mathbf{c}),~-2\le \mathbf{c} \le 2
\end{align}
Note that the prior distribution of the latent variable, $\mathbf{c}$, in EISGAN was set as a standard Gaussian distribution.
Therefore, the range of the values from $-2$ to 2 represents the 95\% confidence interval.
Fig.~\ref{fig:GeneratedEIS} shows the generated EIS data corresponding to stages 3, 5, and 6. 
For stage 5, two latent variables $c_1$ and $c_2$ are selected, which were highly correlated with the capacity changes according to the cycle.
The EIS changes as the latent variables varied from $-2$ to $2$ are shown in Fig.~\ref{fig:GeneratedEIS}~(a) and (b).
Fig.~\ref{fig:GeneratedEIS}~(a) shows that the EIS curve changes when $c_1$ varies, and Fig.~\ref{fig:GeneratedEIS}~(b) shows that the semicircle is divided into two when $c_2$ varies.
Both these changes in the EIS curve can be considered to agree with the battery aging-related changes in EIS data reported in previous studies.
Fig.~\ref{fig:GeneratedEIS}~(c) and (d) show the EIS changes according to $c_1$ for stages 3 and 6, respectively.
In such EIS data measured with DC, where fluctuations occur in the low-frequency region, changes in the degree of bending or the curvature of the low-frequency impedance were observed in both stages 3 and 6.
As a result, we confirmed that the trained generator $G$ of the EISGAN could construct the shape of the EIS data according to the latent variables for each stage.

\begin{figure}[t]
    \centering
    \includegraphics[width=.8\columnwidth]{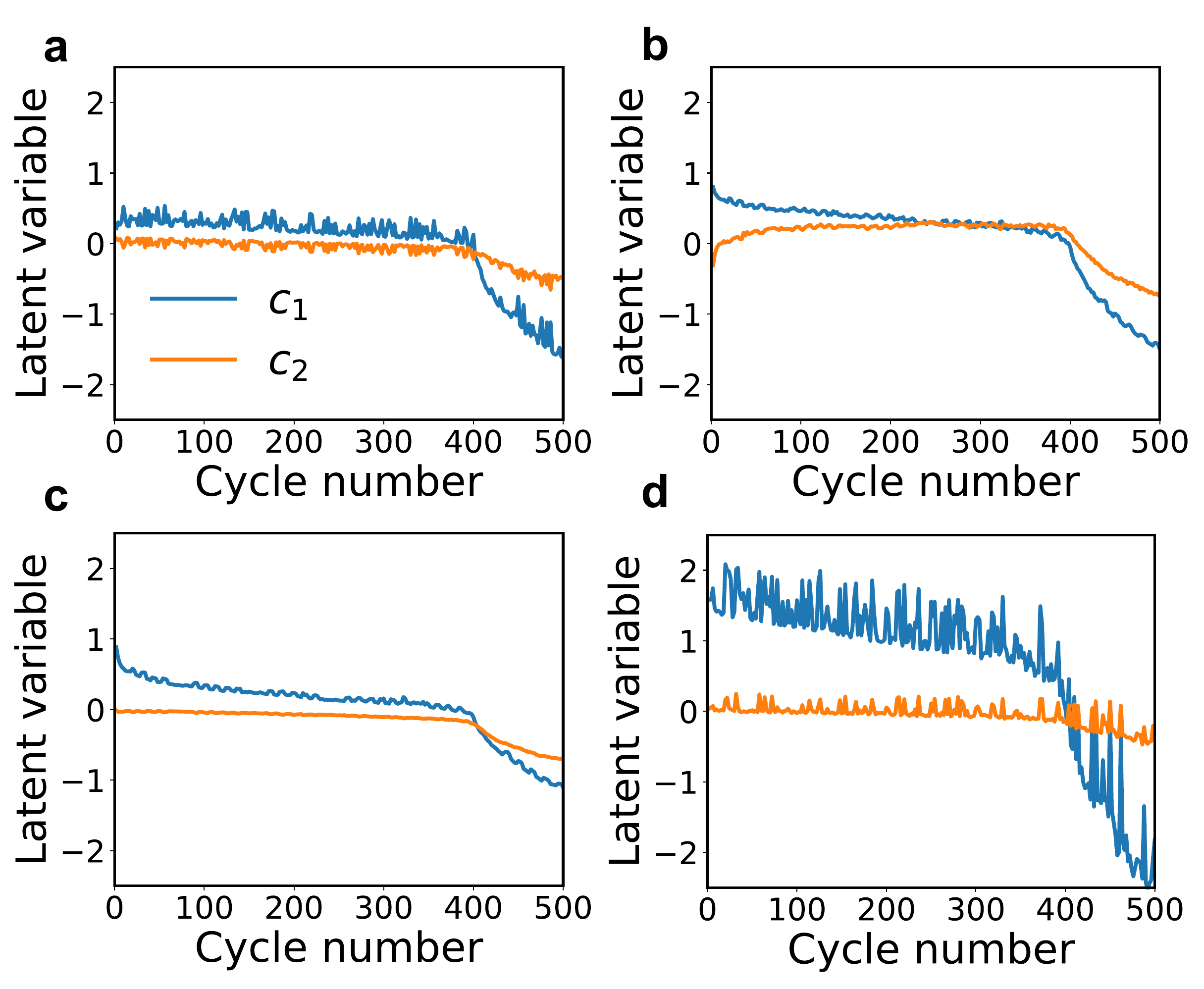}
    \caption{Extracted latent variables of 25C05 for (a) stage 3, (b) stage 4, (c) stage 5, and (d) stage 6. The two latent variables $c_1$ and $c_2$ are selected based on the correlation with the capacity and arranged in decreasing order with the cycles.}
    \label{fig:Latents}
\end{figure}

Changes in the latent variables over cycles, which were extracted from EIS measurements, for stages 3, 4, 5, and 6 of cell 25C05 are presented in Fig.~\ref{fig:Latents}.
Among the nine latent variables for each stage, two variables ($c_1$ and $c_2$) that had a high correlation with the capacity changes over cycles were selected.
It should be noted that latent variables are arranged to decrease with respect to increasing cycles because their signs are not physically restricted.
Overall, the values of $c_1$ gradually change according to the cycle and then exhibit rapid changes at 400 cycles when the battery capacity drops.
By contrast, the values of $c_2$ remain close to 0 until 400 cycles and then change.
Fig.~\ref{fig:Latents}~(a) and (d) present the significant noise in the latent variables; this noise is generated when EIS data are measured with DC, as in stages 3 and 6.
However, in stage 4, where the EIS data were measured without relaxation, the noise in the latent variables was low, as shown in Fig.~\ref{fig:Latents}~(b).
In stage 5, which is the most stable condition, the noise in the latent variables was small, similar to that in in stage 4, as shown in Fig.~\ref{fig:Latents}~(c).
Although the EISGANs were trained separately for each stage, the changes in latent variables over cycles exhibit similar trends among the stages.
Furthermore, we confirmed that the trends of the latent variables reflect the capacity degradation of the battery, and consequently, capacity estimation can be performed through the latent variables.
Some of the latent variables clearly represent the characteristics associated with the SOH, but no evident physical meaning has been identified.
Therefore, it is necessary to map the latent variables with the physical parameters of LIBs.

\subsection{Capacity estimation}

\begin{figure}[t]
    \centering
    \includegraphics[width=.8\columnwidth]{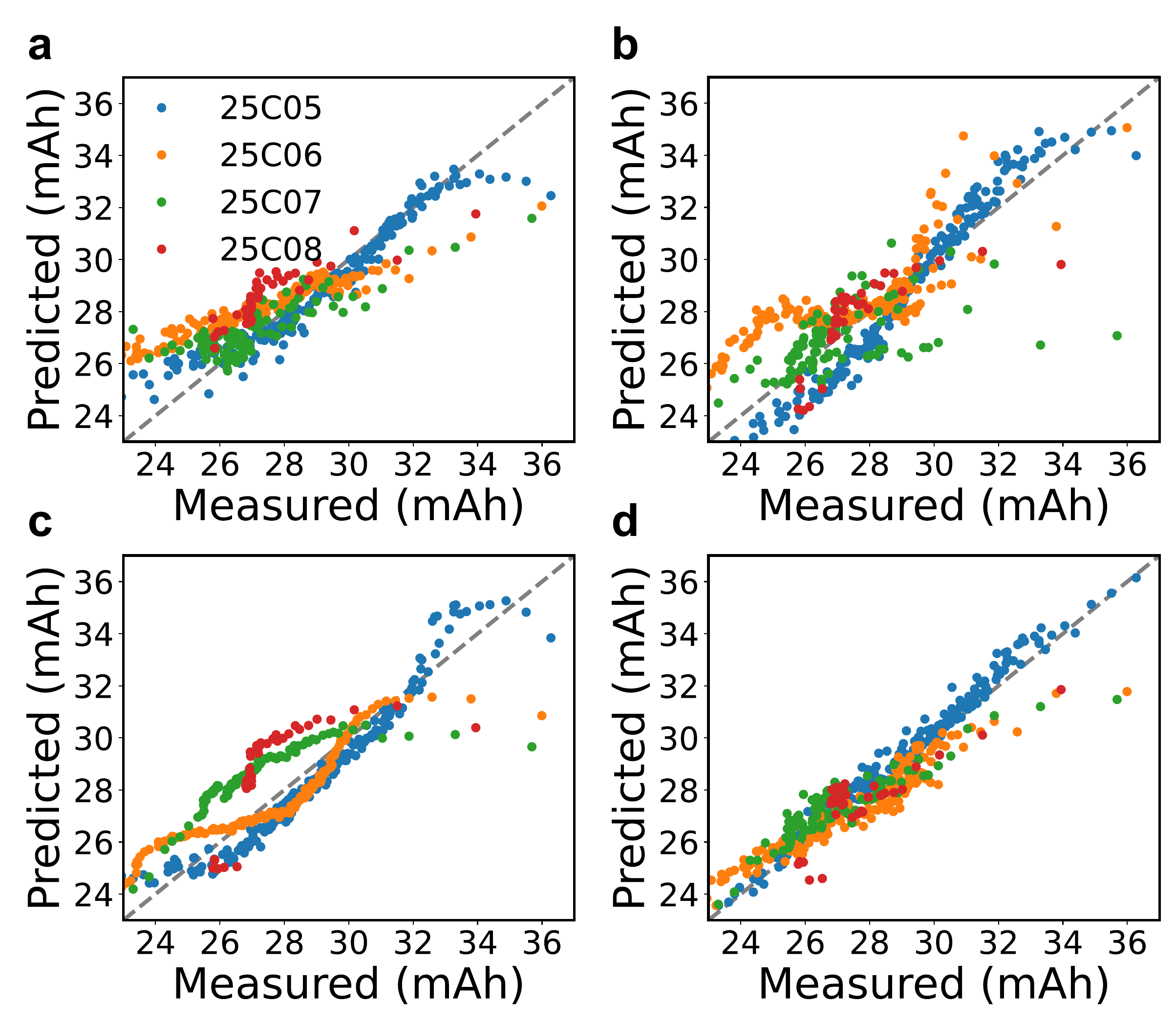}
    \caption{Predicted capacity with respect to the measured capacity of four testing cells using EISGAN for (a) stage 3, (b) stage 4, (c) stage 5, and (d) stage 6.}
    \label{fig:SOH_qq}
\end{figure}

A comparison of the capacity predicted by GPR using the latent variables estimated through EISGAN and the measured capacity for the testing batteries in stages 3--6 is presented in Fig.~\ref{fig:SOH_qq}.
Fig.~\ref{fig:SOH_qq}~(a) shows the results for stage 3, where the MAE and RMSE of all the cells are 0.7879 mAh and 1.093194 mAh, respectively.
The noise involved in the capacity estimation is attributed to the fluctuations in the latent variables.
By contrast, the results for stage 4 indicate that the predicted capacity is scattered, despite the low noise in the latent variables, as shown in Fig.~\ref{fig:SOH_qq}~(b).
The MAE and RMSE are 1.0015 mAh and 1.3588 mAh, respectively, which are larger than those in stage 3.
Nevertheless, the EISGAN appropriately predicted the trend of capacity degradation for each cell.
Fig.~\ref{fig:SOH_qq}~(c) shows the results for stage 5 under the most stable conditions, where the noise in the predicted capacity is relatively smaller than those for stage 3 and 4.
However, for 25C07 and 25C08, the EISGAN tends to underestimate capacity slightly.
The MAE and RMSE for stage 5 are 1.0018 mAh and 1.2307 mAh, respectively, which are similar to those for stage 4.
The predicted capacity was in good agreement with the measured capacity in the case of stage 6, as shown in Fig.~\ref{fig:SOH_qq}~(d), where the low-frequency fluctuations are significantly large.
The MAE and RMSE are 0.5790 mAh and 0.7566 mAh, respectively, implying that the EIS measurements for stage 6 possess regularity, even though they do not exhibit a typical EIS curve.
Overall, we confirmed that adequate capacity predictions with lower than 1.002 mAh of MAE and 1.359 mAh of RMSE at any stage were possible with GPR using the latent variables extracted through the EISGAN.

For the comparison with the proposed EISGAN, capacity prediction through GPR by directly using EIS data, referred to EIS-Capacity GPR~\cite{zhang2020identifying}, was also performed.
The results of capacity estimation through EISGAN and EIS-Capacity GPR for each stage and cell are listed in Table~\ref{tab:results}.
As described in section~\ref{sec:implemenddetail}, the prediction results were evaluated using MAE, RMSE, and $R^2$.
In the case of EIS-Capacity GPR, the deviation between each metric is large for each cell within one stage, whereas for EISGAN, the deviation is relatively small.
For example, in stage 3, minimum and maximum MAEs among four cells for EIS-Capacity GPR are 1.2741 mAh and 5.1149 mAh, respectively, while those of EISGAN are 0.4763 mAh and 1.2668 mAh, respectively.
Besides, in the case of EISGAN, the deviation of each metric between stages is also relatively small, where the minimum and maximum MAEs in stages 3, 4, 5, and 6 of 25C05 are 0.3806 mAh and 0.8647 mAh, respectively.
However, for EIS-Capacity GPR, the minimum and maximum MAEs are 0.4282 mAh and 3.0952 mAh, respectively.
Note that $R^2$ can be a negative value, meaning the capacity prediction has a larger residual error than the horizontal prediction line over the cycle, indicating that the prediction is not suitable.
This phenomenon appears only in one case of 25C08 in stage 5 for EISGAN, but in eight cases for EIS-Capacity GPR, implying a limitation in generalization when EIS measurements are directly entered into GPR.
Overall, considering the value of $R^2$, both EISGAN and EIS-Capacity GPR predict the capacity degradation trend better for 25C05 than other cells, probably because the EIS measurements of 25C05 contain more characteristics indicating the capacity than those of other cells.
Finally, the MAE and RMSE are less than 2 mAh for all stages and cells for the EISGAN, which indicates that the EISGAN shows the robust and reliable estimation of the capacity.

\begin{table}[t]
\centering
\caption{Capacity estimations using EISGAN and EIS-Capacity GPR~\cite{zhang2020identifying} for all testing data.}\label{tab:results}
\resizebox{.65\columnwidth}{!}{\begin{tabular}{lccccr}
\toprule
Method  & {Stage}        & {Cell} & {MAE (mAh)}         & {RMSE (mAh)}        & {$R^2$}                \\
\midrule
\textbf{EISGAN}           & 3                     & 25C05         & 0.4763                     & 0.7772                     & 0.8622                     \\
                 &                       & 25C06         & 0.9679                     & 1.2637                     & 0.5976                     \\
                 &                       & 25C07         & 0.8640                     & 1.1602                     & 0.6175                     \\
                 &                       & 25C08         & 1.2668                     & 1.3524                     & 0.2675                     \\
                 \rule{0pt}{3ex}
                 & 4                     & 25C05         & 0.8647                     & 1.0496                     & 0.7488                     \\
                 &                       & 25C06         & 1.1557                     & 1.5206                     & 0.4174                     \\
                 &                       & 25C07         & 1.0265                     & 1.6138                     & 0.2600                     \\
                 &                       & 25C08         & 0.9154                     & 1.1595                     & 0.4616                     \\
                 \rule{0pt}{3ex}
                 & 5                     & 25C05         & 0.7256                     & 0.8569                     & 0.8326                     \\
                 &                       & 25C06         & 0.7127                     & 0.9114                     & 0.7907                     \\
                 &                       & 25C07         & 1.6835                     & 1.8063                     & 0.0729                     \\
                 &                       & 25C08         & 1.7357                     & 1.8676                     & -0.3970                    \\
                 \rule{0pt}{3ex}
                 & 6                     & 25C05         & 0.3806                     & 0.4883                     & 0.9456                     \\
                 &                       & 25C06         & 0.6377                     & 0.8267                     & 0.8278                     \\
                 &                       & 25C07         & 0.7602                     & 0.9489                     & 0.7441                     \\
                 &                       & 25C08         & 0.7765                     & 0.8910                     & 0.6820                     \\
                 \rule{0pt}{3ex}
                 & *7                     & 25C07         & 0.7248                     & 0.9537                     & 0.7416                     \\
                 &                       & 25C08         & 0.6392                     & 0.7952                     & 0.7467                     \\
\midrule
EIS-Capacity GPR~\cite{zhang2020identifying} & 3                     & 25C05         & 3.0952                     & 3.2382                     & -1.3913                    \\
                 &                       & 25C06         & 1.2741                     & 1.4734                     & 0.4530                     \\
                 &                       & 25C07         & 5.1149                     & 5.1361                     & -6.4963                    \\
                 &                       & 25C08         & 4.7250                     & 4.8994                     & -8.6981                    \\
                 \rule{0pt}{3ex}
                 & 4                     & 25C05         & 1.2960                     & 1.5093                     & 0.4805                     \\
                 &                       & 25C06         & 2.9443                     & 3.4186                     & -1.9446                    \\
                 &                       & 25C07         & 1.7625                     & 2.0091                     & -0.1471                    \\
                 &                       & 25C08         & 1.3408                     & 1.4104                     & 0.1963                     \\
                 \rule{0pt}{3ex}
                 & 5                     & 25C05         & 0.4282                     & 0.5797                     & 0.9234                     \\
                 &                       & 25C06         & 2.3790                     & 2.9320                     & -1.1660                    \\
                 &                       & 25C07         & 2.7055                     & 3.2278                     & -1.9607                    \\
                 &                       & 25C08         & 0.5760                     & 0.6689                     & 0.8192                     \\
                 \rule{0pt}{3ex}
                 & 6                     & 25C05         & 1.2842                     & 1.5750                     & 0.4343                     \\
                 &                       & 25C06         & 0.5757                     & 0.6398                     & 0.8969                     \\
                 &                       & 25C07         & 0.8853                     & 0.9552                     & 0.7407                     \\
                 &                       & 25C08         & 0.6750                     & 1.1041                     & 0.5075                     \\
                 \rule{0pt}{3ex}
                 & *7                     & 25C07         & 1.2224                     & 2.2332                     & -0.4172                    \\
                 &                       & 25C08         & 0.9999                     & 1.5497                     & 0.0297                     \\
\bottomrule
\multicolumn{6}{l}{*Note that 25C04 and 25C06 were used to train the models for stage 7.}
\end{tabular}}
\end{table}

\begin{figure}[t]
    \centering
    \includegraphics[width=.8\columnwidth]{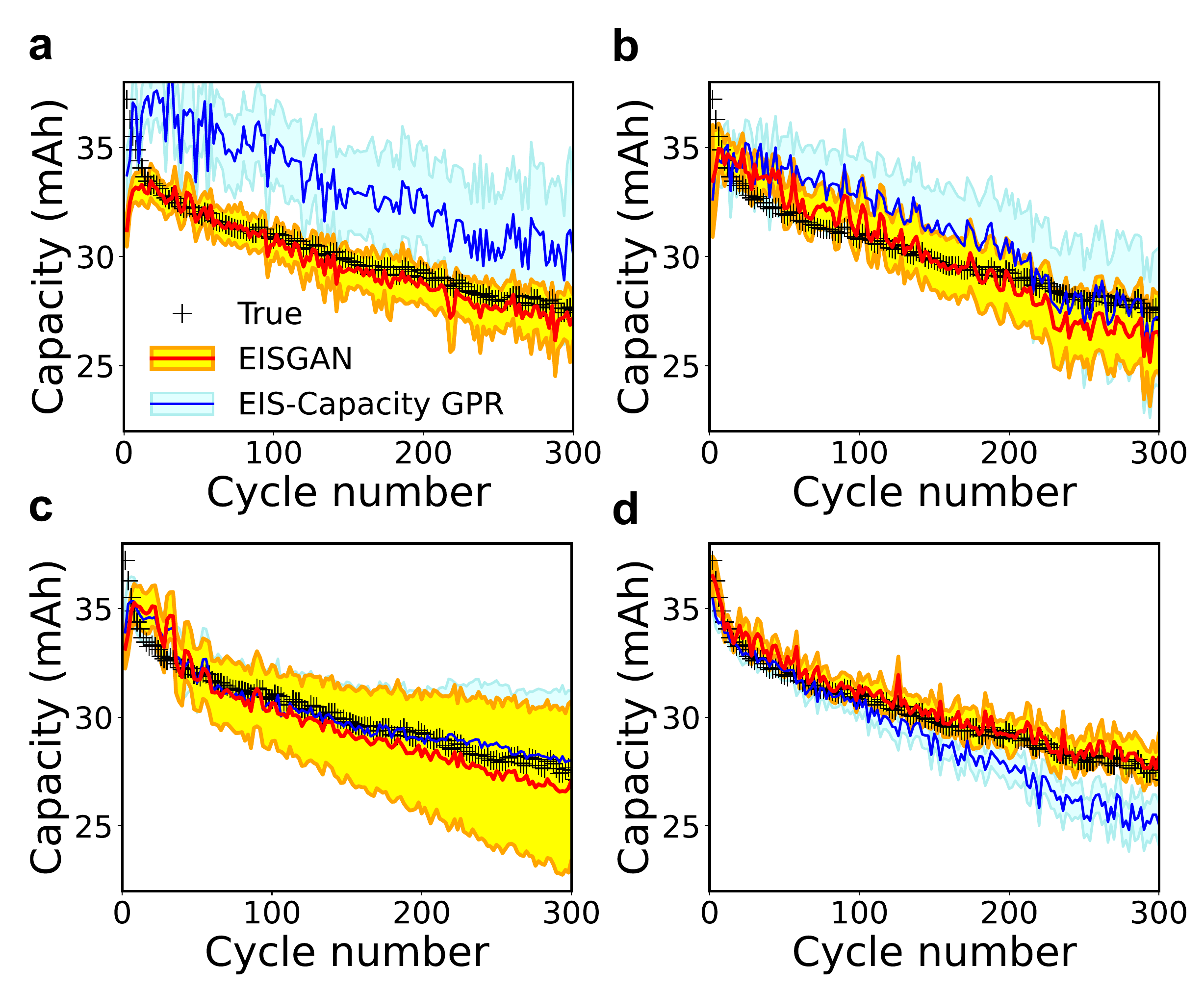}
    \caption{Capacity estimations for 25C05 using EISGAN, as compared to those using EIS-Capacity GPR for (a) stage 3, (b) stage 4, (c) stage 5, and (d) stage 6. The shaded region indicates a standard deviation of $\pm1$.}
    \label{fig:SOH_Comparison}
\end{figure}

Capacity and confidence interval estimations of 25C05 using EISGAN and EIS-Capacity GPR compared to measured capacity according to cycles for stages 3--6 are shown in Fig.~\ref{fig:SOH_Comparison}.
In stage 3, shown in Fig.~\ref{fig:SOH_Comparison}~(a), where the EIS data were measured with DC during charging, the EISGAN shows a stable prediction while the EIS-Capacity GPR tends to overestimate the capacity.
Furthermore, the confidence interval of EISGAN is narrower than that of EIS-Capacity GPR, which indicates the reliable prediction of the EISGAN.
In stage 4, where the EIS data were measured right after charging, the EISGAN reliably estimates the capacity, but EIS-Capacity GPR slightly overestimates the capacity until early 200 cycles, as shown in Fig.~\ref{fig:SOH_Comparison}~(b).
However, after 200 cycles, the EISGAN underestimate the capacity, while the predicted confidence interval still encompasses the measured capacity values.
Fig.~\ref{fig:SOH_Comparison}~(c) shows the results in stage 5, the most stable condition with 15 minutes of rest after charging, where both EISGAN and EIS-Capacity GPR show adequate predictive performance.
The confidence interval for both EISGAN and EIS-Capacity GPR is wider in stage 5 than in other stages, and the confidence interval increases as the cycle increases.
Note that the EISGAN tends to underestimate capacity in the initial cycles in stages 3, 4, and 5, which also occurs in the case of EIS-Capacity GPR.
In stage 6, the EISGAN reliably estimates the capacity though the intense noises occur in latent variables over cycles, as shown in Fig.~\ref{fig:SOH_Comparison}~(d).
On the other hand, the EIS-Capacity GPR underestimates the capacity as the cycle increases.
The results indicate that the EISGAN is insensitive to the presence of DC and relaxation when compared to the EIS-Capacity GPR, and reliably estimates the capacity.
The capacity estimations for 25C06, 25C07, and 25C08 are shown in Fig.~\ref{fig:EISGAN_Others} in Appendix.

\subsection{Robustness evaluation}

\begin{figure}[t]
    \centering
    \includegraphics[width=.8\columnwidth]{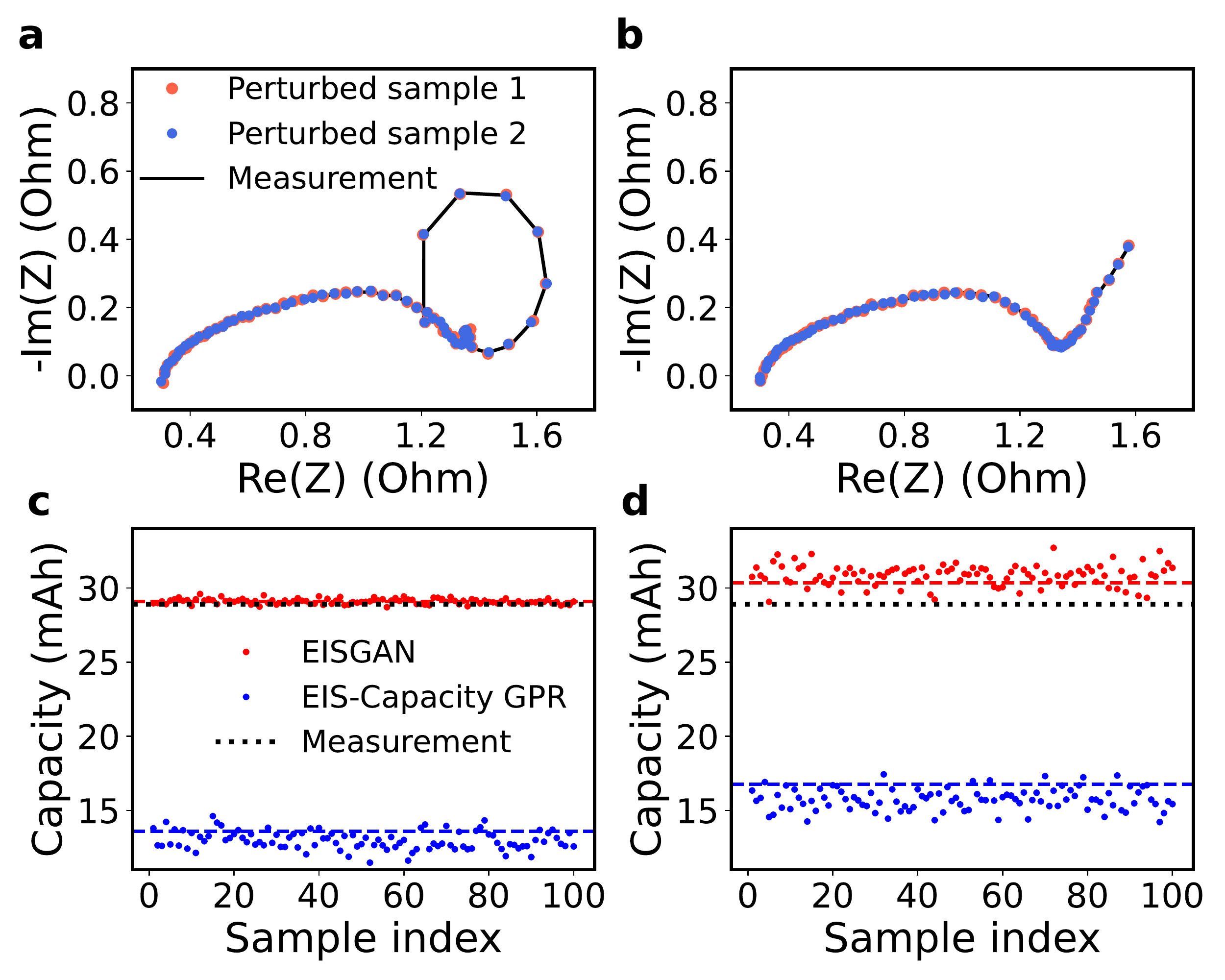}
    \caption{Perturbed samples of EIS data of 25C06 with $\sigma=0.003$ for (a) stage 3 and (b) stage 5, and estimated capacity for 100 samples for (c) stage 3 and (d) stage 5. Dotted colored lines represent the estimated capacity for measured EIS data using EISGAN and EIS-Capacity GPR.}
    \label{fig:perturbed_EIS}
\end{figure}

\red{
To analyze the reason why the EISGAN yielded better results than those of the EIS-Capacity GPR, a perturbation approach was performed on the EIS data to evaluate the robustness of the models.
Perturbations were applied in the form of adding Gaussian noise $\Delta Z$, with zero mean and variance $\sigma^2$, to the measured EIS curve $Z$ based on the suggestions reported in \cite{agarwal1995application}.
The perturbed EIS curve $Z'$ can be presented as follows:
\begin{align}
    Z'=Z+\Delta Z,~\text{where}~\Delta Z\sim N(0, \sigma^2).
\end{align}
Using the perturbed EIS data, the capacity was estimated via the EISGAN and EIS-Capacity GPR, and deviations from the estimated capacity using the measured EIS data were analyzed.
Among the test cells, the EIS data from stages 3, 4, 5, and 6 of the 100th cycle of 25C06 were selected, and the data were perturbed with $\sigma=0.001, 0.003$, and $0.005$.
Figs.~\ref{fig:perturbed_EIS}~(a) and (b) show the two perturbed samples in stages 3 and 5, respectively, when $\sigma = 0.003$.
The black solid lines represent the measured EIS curves.
Figs.~\ref{fig:perturbed_EIS}~(c) and (d) depict the capacity estimation results of the models for 100 perturbed samples of stages 3 and 5, respectively.
The black dotted lines indicate the measured capacity, while the red and blue dotted lines indicate the capacity estimated via the EISGAN and EIS-Capacity GPR using a measured EIS curve, respectively.
In the case of stage 3, the capacity estimated using the EISGAN for the perturbed EIS data showed slight deviations compared to that of the EIS-Capacity GPR.
However, in the case of stage 5, the deviations in the estimated capacity for the perturbed EIS data of the two models were similar.
}

\begin{figure}[t]
    \centering
    \includegraphics[width=.5\columnwidth]{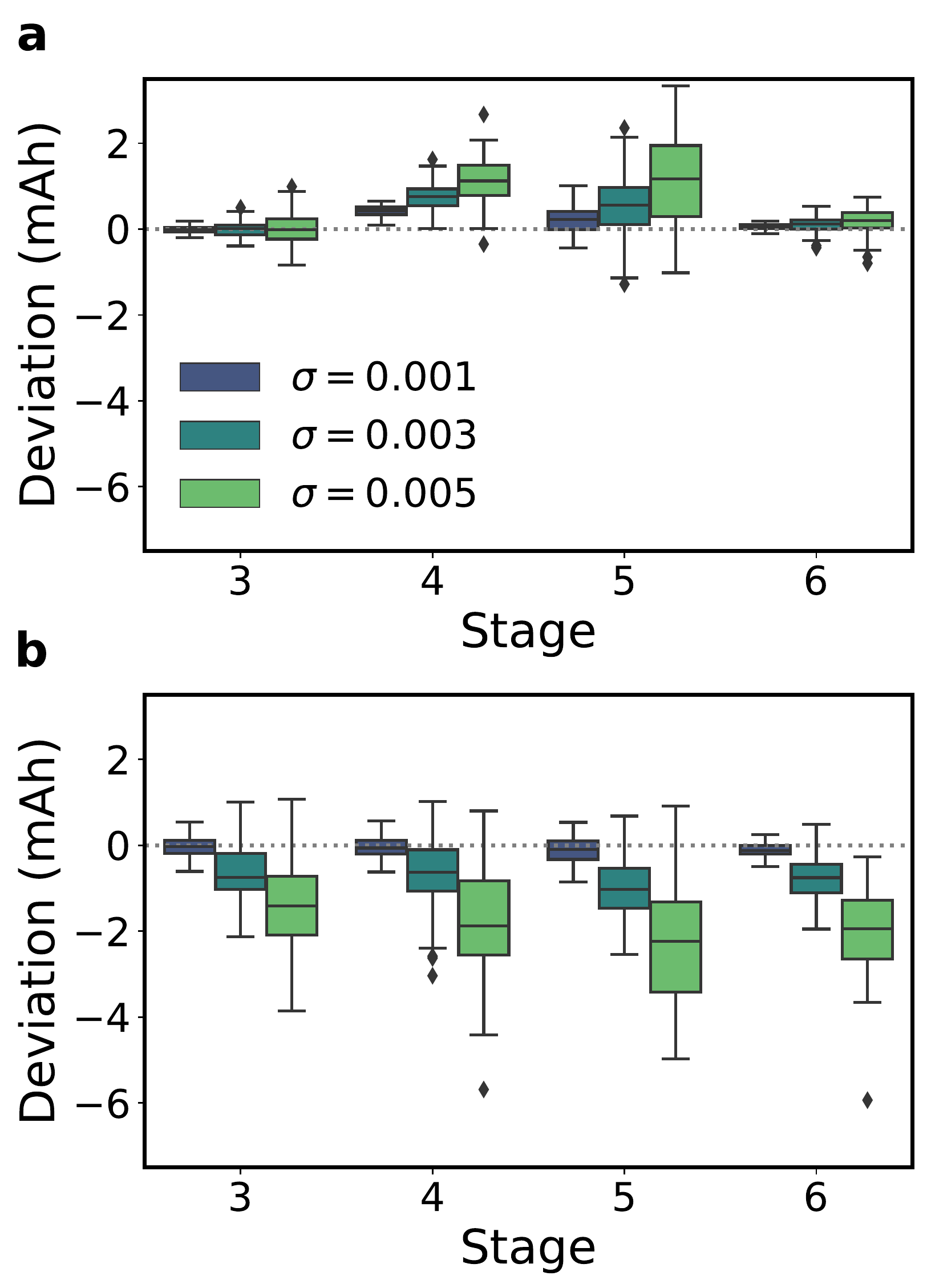}
    \caption{Deviation in capacity estimations between measured and perturbed data for different stages and $\sigma$s using (a) EISGAN and (b) EIS-Capacity GPR.}
    \label{fig:perturbed_sumamry}
\end{figure}

\red{
For each stage, the deviations in capacity for the 100 perturbed EIS data, estimated via the EISGAN and EIS-Capacity GPR according to $\sigma$, are summarized in the forms of box plots in Figs.~\ref{fig:perturbed_sumamry}~(a) and (b), respectively.
In the box plot, the central line indicates the median, the edges of the box show the 25th and 75th percentile values, the whiskers extend to the most extreme data points that are not considered outliers, and the outliers are plotted individually with `$\blacklozenge$'.
In both models, as $\sigma$ increased, the median as well as the range of the deviations tended to increase.
In the case of the EISGAN in Figs.~\ref{fig:perturbed_sumamry}~(a), the median of deviations in stages 3 and 6 did not change significantly from 0, and only the range increased.
However, in stages 4 and 5, both the median and range increased; particularly, the range of deviations increased significantly in stage 5.
This indicates that the perturbation applied to the EIS curve of the cell, having been sufficiently rested after charging, greatly affected the performance of the model.
Meanwhile, in the case of the EIS-Capacity GPR in Figs.~\ref{fig:perturbed_sumamry}~(b), the median and range of deviations increased for all stages.
Moreover, the increased median and range of deviations were similar or larger in all cases than those observed for the EISGAN under the same conditions.
The variables extracted through the EISGAN can sufficiently express the characteristics of the observational data with a small degree of freedom, and the effect of the observation errors can be reduced through this feature-extraction process~\cite{ren2018remaining, guo2019data}.
Moreover, the median of deviations was positive for the EISGAN and negative for the EIS-Capacity GPR, because the EISGAN and EIS-Capacity GPR overestimated and underestimated the capacity, respectively, while using the measured EIS data.
These results indicate that it is more advantageous in terms of robustness against the data noise, to estimate the capacity using the latent variables extracted through the EISGAN, rather than directly entering the EIS data.
}


\section{Conclusions}\label{sec5:conclusion}

This paper proposes the EISGAN for extracting latent variables that can reliably represent the characteristics of LIBs from EIS data, even when measured with DC and without relaxation.
The proposed method was validated under various conditions according to the relaxation and DC during charging and discharging.
Using the EISGAN, changes in the generated EIS data were examined by varying each latent variable from $-2$ to $2$.
We confirmed that reliable EIS data could be generated for each stage.
Furthermore, changes in the latent variables of the EIS measurements over cycles were investigated as capacity degradation progressed, and the trends of the latent variables reflected the capacity degradation.
The extracted latent variables were used to estimate the capacity of the batteries through GPR, and the simulation results indicate robust and reliable capacity estimations.
The MAE and RMSE of all the testing cells were below 2 mAh for all stages, with slight variations within and among stages.
The performance of the EISGAN was compared with that of directly obtaining capacity predictions from EIS measurements, which is referred to as EIS-Capacity GPR.
It was found that the EISGAN outperforms EIS-Capacity GPR when using EIS data measured under various conditions.
We demonstrate that the EISGAN can extract meaningful representations from EIS measurements that are difficult to address using circuit models.
Despite the contribution, the shortcoming of the proposed method is that the physical meaning of the latent variables must be inferred through changes in the generated EIS data.
In future studies, the EISGAN will be compared with circuit models and other physical or data-driven models for the SOH estimation of LIBs.
Moreover, a comparative performance evaluation of the EISGAN with other unsupervised models will also be conducted.
In the case of a BMS-equipped system, since the SOC value during the charging and discharging phases is known, the SOH estimation can be performed practically with a separate EISGAN model for each stage.
However, additional experiments and related sensitivity studies must be performed in future studies to obtain more practical SOH estimations at any stage.
Furthermore, the proposed method can be applied using the capacity, terminal voltage, and current-voltage curves along with the EIS data in the future studies.


\section*{Acknowledgment}

%

This work was supported by the National Research Foundation of Korea (NRF) grant funded by the Korean government (Ministry of Science and ICT) (NRF-2017R1E1A1A0-3070161 and NRF-20151009350), and the National Supercomputing Center with supercomputing resources including technical support (KSC-2020-INO-0056).
{This research was partially supported by the Graduate School of Yonsei University Research Scholarship Grants in 2020.} {Computing resources was also supported by the National IT Industry Promotion Agency (NIPA)}.

\bibliographystyle{elsarticle-num}
\bibliography{bibfile}

\clearpage

\appendix


\section{EIS measurements of eight cells}

\begin{figure}[h!]
    \renewcommand{\thefigure}{A.1}
    \centering
    \includegraphics[width=.75\columnwidth]{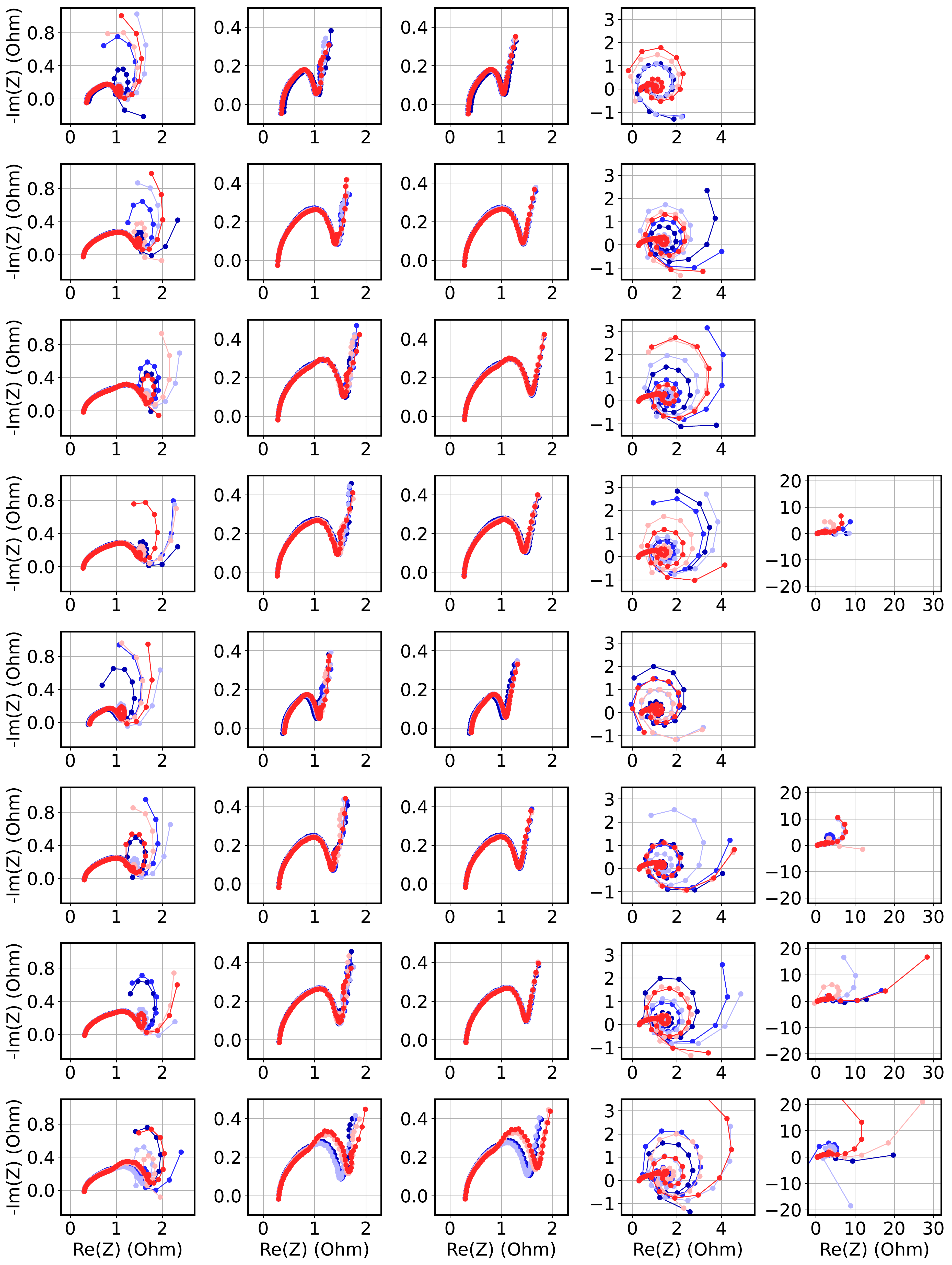}
    \caption{EIS data of cells 25C01--25C08 in rows collected at stages 3--7 in columns over 20, 40, 60, 80, and 100 cycles.}
    \label{fig:EIS_All}
\end{figure}

The EIS measurements of the eight cells used in the study are depicted in Fig.~\ref{fig:EIS_All}.
The first four cells (25C01--25C04) were used to train the EISGAN and the remaining four cells (25C05--25C08) were used to validate the model for stages 3, 4, 5, and 6.
For the stage 7, the cells 25C04 and 25C06 were used to train the model, while the 25C07 and 25C08 were used for validation.

\clearpage

\section{Capacity estimation results}

\begin{figure}[h!]
    \renewcommand{\thefigure}{B.1}
    \centering
    \includegraphics[width=.9\textwidth]{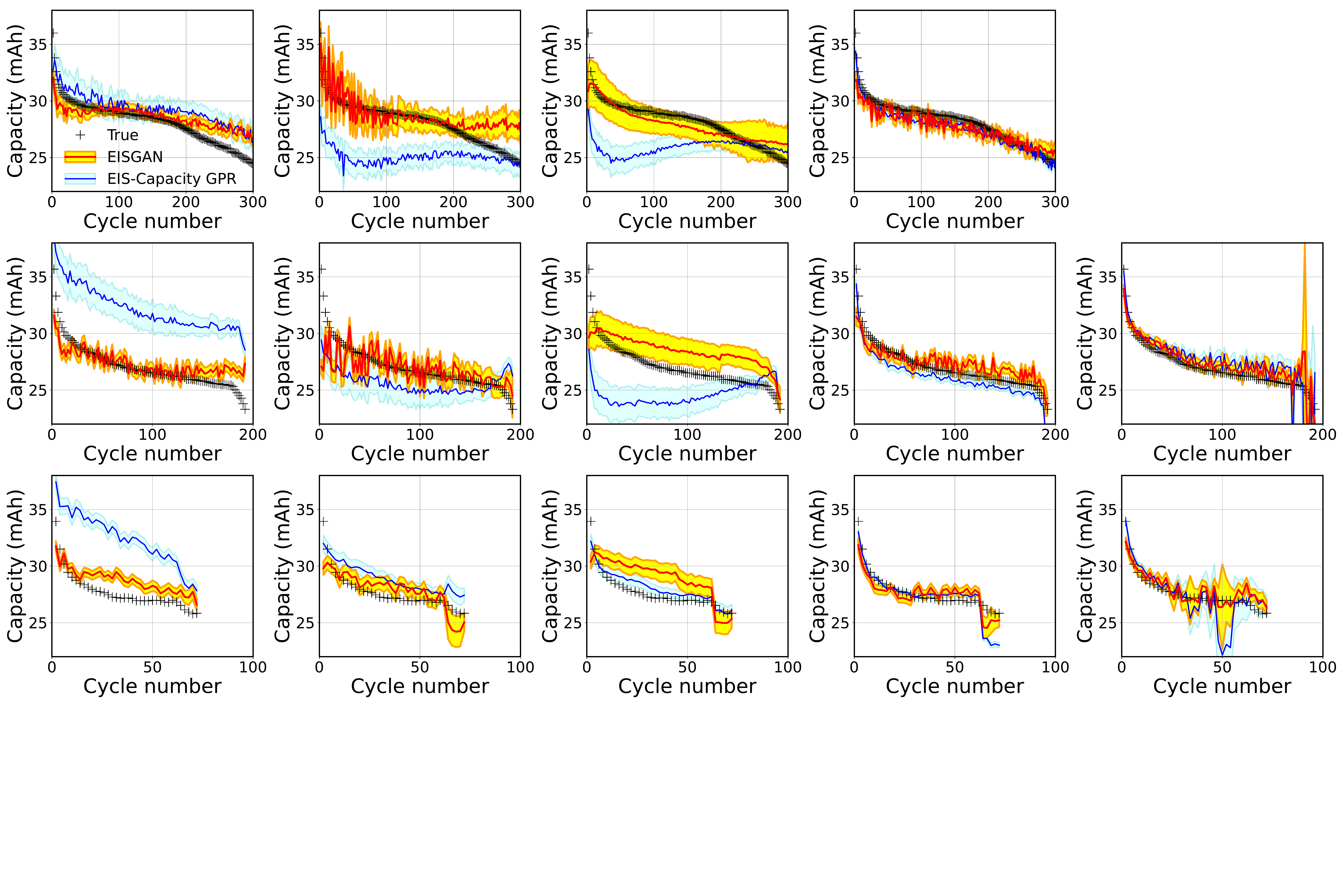} \\
    \vspace{-2cm}
    \caption{Capacity estimations for 25C06--25C08 in rows using EISGAN, as compared to those using EIS-Capacity GPR for stage 3--7 in columns. The shaded region indicates a standard deviation of $\pm1$.}
    \label{fig:EISGAN_Others}
\end{figure}

The capacity estimation result of the remaining three test batteries (25C06, 25C07, and 25C08) are depicted in Fig.~\ref{fig:EISGAN_Others}.
Overall, the estimated capacity of the EISGAN properly follows the trend of the measured capacity correctly, whereas that of the EIS-Capacity GPR shows a relatively large bias.
The results indicate that the EISGAN is insensitive to the presence of DC and relaxation, compared to the EIS-Capacity GPR, and provides a reliable estimation of the capacity.

\end{document}